%% file: Main.tex
\documentclass[sigconf]{acmart}
\AtBeginDocument{%
  }

\setcopyright{acmlicensed}

\copyrightyear{2026}
\acmYear{2026}
\setcopyright{cc}
\setcctype{by}
\acmConference[CHI EA '26]{Extended Abstracts of the 2026 CHI Conference on Human Factors in Computing Systems}{April 13--17, 2026}{Barcelona, Spain}
\acmBooktitle{Extended Abstracts of the 2026 CHI Conference on Human Factors in Computing Systems (CHI EA '26), April 13--17, 2026, Barcelona, Spain}
\acmDOI{10.1145/3772363.3798971}
\acmISBN{979-8-4007-2281-3/2026/04}

\begin{document}

\title{Insights from Farmer-Managed Decentralized Solar Irrigation Systems}

\author{Arnab Paul Choudhury}
\affiliation{%
  \institution{Viksit Labs Foundation}
  \city{Silchar}
  \state{Assam}
  \country{India}}
\email{arnabpchoudhury@viksitlabs.in}

\author{Rahul Rathod}
\authornote{Work done while the author was working at the IWMI-Tata Program of the International Water Management Institute-CGIAR}
\affiliation{%
  \institution{Sustain Plus Energy Foundation}
  \city{Anand}
  \state{Gujarat}
  \country{India}}
\email{rathodrahul911@gmail.com}

\author {Aryan Yadav}
\affiliation{%
  \institution{Viksit Labs Foundation}
  \city{Silchar}
  \state{Assam}
  \country{India}}
\email{aryanyadav@viksitlabs.in}

\renewcommand{\shortauthors}{Paul Choudhury et al.}

\begin{abstract}
  Solar irrigation systems are increasingly deployed in rural regions, yet their distributed and remote deployment makes maintenance challenging for farmers. While formal monitoring processes and applications exist, they often fall short in practice. We present insights from grid-connected solar irrigation schemes that incentivize farmers to feed energy to the grid, focusing on how farmers maintain their systems. We found that farmers face multiple challenges but are also devising strategies, including the appropriation of WhatsApp to share daily generation data with peers and compare performance across installations to identify potential system anomalies. Our findings highlight how messaging platforms function as informal digital infrastructures enabling collective sensemaking around distributed energy systems. We discuss implications for designing agricultural energy technologies that support peer comparison, contextual interpretation, and community-driven maintenance, framing these as a socio-technical platform. Finally, we outline directions for future work integrating such practices with formal monitoring tools and explore their potential to support citizen science initiatives in environmental sensing.
\end{abstract}




\maketitle

\input{1.Introduction}

\input{2.Methods}
\input{3.Findings}
\input{4.Discussions}
\input{5.Conclusions}
\begin{acks}
We thank our participants for sharing their experiences. We also thank the anonymous reviewers for offering valuable insights that helped us in improving this paper.
\end{acks}

\bibliographystyle{ACM-Reference-Format}
\bibliography{references}

\end{document}

%% file: 1.Introduction.tex
\section{Introduction}
Throughout the world, agriculture 4.0 is transforming farming through IoT, AI, robotics, and big data \cite{AHMED2025100848}. Precision agriculture, powered by GNSS, GIS, remote sensing, and drones, enhances resource efficiency \cite{jsan13040039}, while AI platforms support crop monitoring, irrigation, and climate adaptation \cite{PARRALOPEZ2024109412, Ahmadi2025, IIITAres22:online}. In India, government programs as well as agritech startups have been piloting and investing in AI-IoT for pest surveillance, crop quality assessment, and multilingual farmer advisories \cite{AIforagr79:online, Jeevanandam2024, Cabinetn66:online}.

However, smallholder farmers in the Global South face barriers such as limited infrastructure, cost, and lack of training \cite{Bhat2025, Odume2024}. Studies also warn that agricultural technologies may reinforce power imbalances, create health and safety risks, thus necessitating participatory design to protect marginalized stakeholders \cite{10.1145/3313831.3376364, 10.1145/3274418, 10.1145/3432932, 10.1145/3613904.3642263}. Unequal access to digital tools, labor-intensive norms, reliance on traditional practices, and adoption only when necessary, further limit the potential of agri-tech \cite{10.1145/3613904.3642099, 10.1145/3706598.3713643, 10.1145/3610222}.

Alongside digital shifts in agriculture, the global energy sector is rapidly transitioning to renewables. Falling renewable technology costs and ambitious climate targets have driven global installed power and energy transition in recent years \cite{IEA2024Renewables, IRENA2024RenewableStats}. In India, renewable capacity surpassed 200 GW in 2024, with a 2030 target of 500 GW \cite{PIB2024RenewableMilestone}, and solar energy has become a key focus, supported by national and state programs \cite{MSEDCL_iSMART2025, Deshpande2025MSEDCL, GujaratEPD_AgricultureSchemes2025, GujaratSolarPolicy2021}. However, challenges remain in scaling renewables, including curtailment, storage, and maintenance issues \cite{IPAG_DRE_2024, TOI2025_RajasthanRenewableEnergy, Raizada2025_IndiaEnergyTransition}. Similarly, many agricultural digitization initiatives fail when they treat farmers as isolated, rational actors, overlooking historical and political contexts, creating perceptions of opaque, low-value AI systems \cite{10.1145/3706598.3714250, 10.1145/3678884.3682046}, and prioritizing short-term gains over long-term sustainability \cite{10.1145/3359136}.

Agriculture sits at the convergence of digital and energy transformations. Integrating solar technologies, such as photovoltaic (PV) irrigation pumps, solar thermal systems, and agrivoltaics, offers sustainable solutions for water-energy-food nexus challenges \cite{agronomy14122791, DOE_Agrivoltaics, Gilchrist2024}. Agrivoltaics, in particular, can generate renewable energy while improving water-use efficiency and crop yields, positively impacting the Water-Energy-Food-Ecosystem (WEFE) nexus \cite{Macknick2019_Agrivoltaics_NREL, UNEPDHI2024_WEFE_Nexus, UNECE_WEFE_Nexus}. In India, several states have piloted solar irrigation pumps \cite{Shah2024_BiharSolarPumps, Shah2016_SolarCrop, Durga2021_SuryaRaithaEPW, TOI_HaryanaSolarGrowth2025, TOI_AP_FarmSolarisation2025}, yet high costs, technical complexity, and policy gaps limit broader adoption \cite{agronomy14122791}. Climate uncertainty further affects sustainability, disproportionately impacting smallholders by influencing irrigation, yields, and livelihoods \cite{ChoquetteLevy2025, Mitter2024, Jatav2024, Thein2025}.

Against this backdrop, our study explores an innovative way through which farmers themselves tackled monitoring challenges by appropriating everyday digital tools. Specifically, we examine the experiences of farmers in a 25-year large-scale pilot run by the Government of Gujarat, India, Suryashakti Kisan Yojana (SKY), i.e., Solar-Powered Farm Scheme, and the initial pilot which influenced the scheme \cite{Shah2016_SolarCrop}, where farmers can utilize Solar irrigation Pumps (SIPs) for irrigation and also export surplus energy to the grid \cite{TheCasef55:online}. We document how farmers informally repurposed WhatsApp groups to collectively monitor their SIPs, compare energy generation, and detect anomalies such as dust accumulation, wiring faults, shading, and more. This grassroots innovation reveals how digital infrastructures not originally designed for energy infrastructure maintenance can evolve into community-driven systems of support, filling gaps left by formal applications. These insights into digital technology’s role in sustaining decentralized solar systems were guided by the research question:

\textbf{RQ:} \textit{How do farmers manage and maintain geographically dispersed Solar Irrigation Pumps? }

Through ethnographic engagement, our findings reveal that digital infrastructures and community participation are crucial components for SIP sustenance. Farmers’ WhatsApp groups exemplify how community practices can embed sustainability and reliability into solar energy infrastructures. This work contributes to HCI, CSCW, and ICTD literature in the following ways:

\begin{itemize}
    \item \textbf{An empirical account of grassroots digital innovation:} We document farmers’ creative use of WhatsApp, a ubiquitous digital infrastructure to sustain and monitor decentralized energy infrastructure, extending CSCW and ICTD discussions of informal digital infrastructures.
    \item \textbf{Implications on design and sustainability:} We discuss how incentivized energy export and digital participation can offer new pathways for participatory governance, resilient infrastructure design, and potentially support citizen-science climate sensing grids in environmental sensing.
\end{itemize}
In sum, this paper frames farmer-managed solar irrigation not only as a technological intervention but as a socio-technical platform, one that entwines everyday digital practices and local collaboration to advance both sustainability and equity in rural contexts.

%% file: 2.Methods.tex
\section{Methodology}
The following sub-sections detail the data sources, methods, and the geographical area of the study. We first describe the region of interest, followed by detailing the data collection strategy and analysis methods used. We followed the "immersion/crystallization" approach, which guided our multi-year engagement with the participants \cite{crabtree2023doing}. The process was iterative, involving multiple stages of data gathering and analysis while remaining reflexive throughout the process. We took verbal informed consent from our participants in the study during our interactions, primary data collection, compilation, and took all measures on our behalf to ensure data is de-identified and their privacy is retained.

\subsection{Geographical Area and Context}
The study was conducted in Gujarat, India’s westernmost state, covering about 196,000 km² along the Arabian Sea and bordering Rajasthan, Madhya Pradesh, and Maharashtra. The state spans diverse terrains, from the arid Rann of Kutch and semi-arid plains to low hills and fertile southern river regions. Gujarat has a tropical monsoon to semi-arid climate, with over 95\% of annual rainfall concentrated in the roughly 15-week southwest monsoon, particularly in the Saurashtra region
\cite{GujaratH66:online, JG_Hirapara}.

\subsection{Participant Recruitment \& Data Collection}
\begin{figure}[htbp]
    \centering
    \includegraphics[width=1\linewidth]{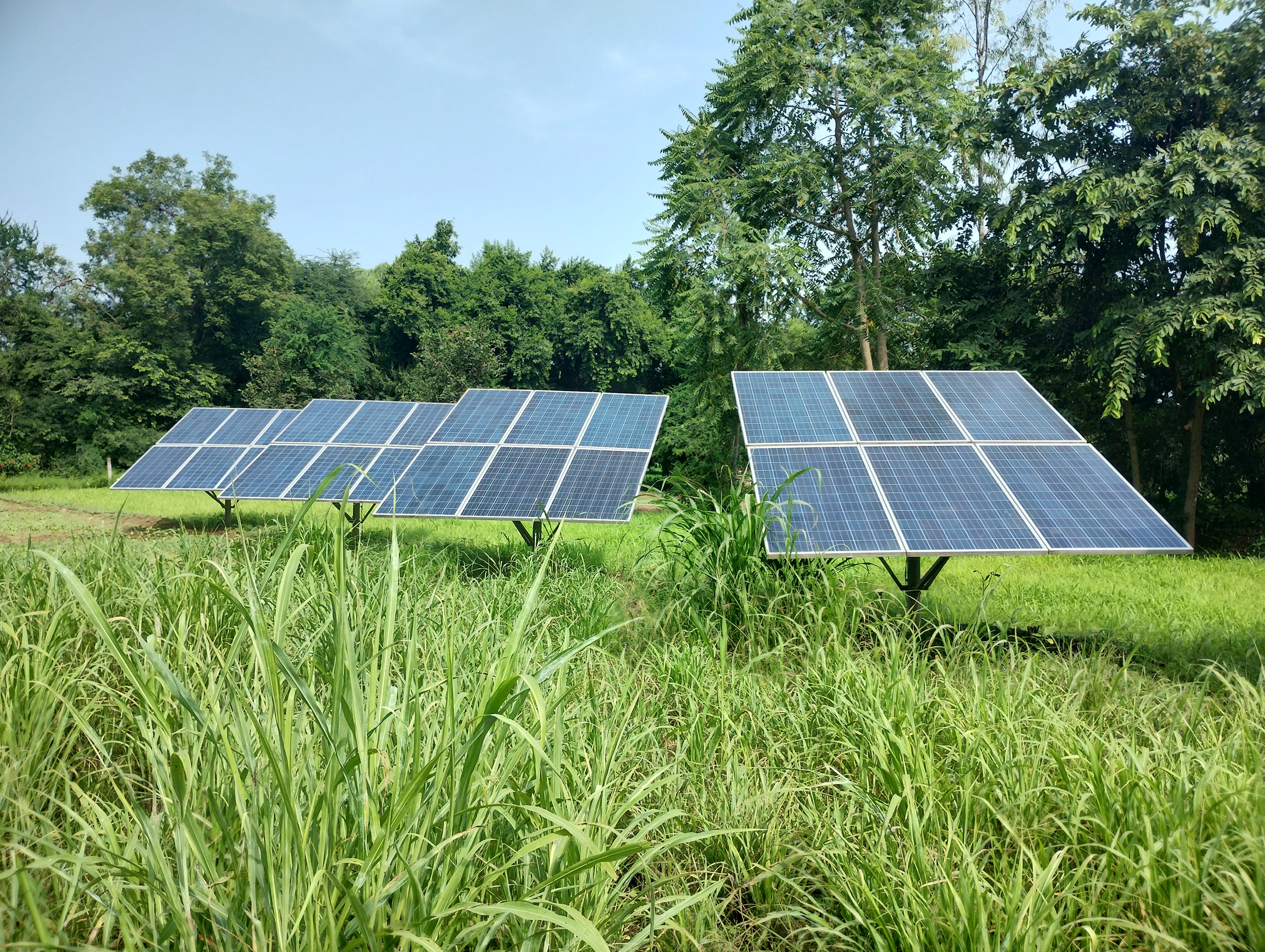}
    \caption{Representative image of a Solar Irrigation Pump installation}
    \label{fig:field}
    \Description{The photograph shows four separate ground-mounted solar panel structures arranged in a staggered line across a grassy field. Each structure holds an array of six dark blue solar panels, configured in two rows of three, all tilted upward to face the sun. The foreground is filled with tall, thick green grass that partially hides the bottom of the metal supports. A tall leafy plant is reaching up to visually overlap and partially obscure the bottom left corner of the frontmost array. Behind the solar panels, a dense, continuous line of lush green trees forms a natural backdrop under a clear, pale blue sky, indicating a bright, sunny day.}
\end{figure}

This study reports on practices by two cohorts of farmers, C1 and C2, connected to two electricity feeders located in the Central Gujarat region. C1 and C2 have 9 and 11 farmers, respectively. While both C1 and C2 worked under a similar financial energy export model, C2 farmers are from one of the 93 feeders participating in the Government of Gujarat’s Suryashakti Kisan Yojana (SKY), that is, Solar-Powered Farmer Scheme (SPFS). C1, on the other hand, was the initial pilot that demonstrated Solar Power as a Remunerative Crop (SPaRC), inspiring the large-scale SPFS scheme and others like the Government of India-backed PM-KUSUM, Component-C (IPS) \cite{TheCasef55:online}. As a result, C1 farmers are not part of the SPFS scheme. Under the solar schemes, farmers use solar power for irrigation as needed, while surplus energy is fed into the grid in exchange for financial compensation. Launched in 2018, SPFS is a 25-year large-scale pilot promoting grid-connected solar irrigation pumps across Gujarat, India \cite{TheCasef55:online, GridConn12:online, UGVCL87:online}. Finally, all C1 farmers were from the same village and organized themselves under a registered cooperative. Similarly, C2 farmers also lived in close proximity to each other; however, they did not form any registered entity.
To support farmers’ understanding of their energy generation, consumption, and evacuation, the Government of Gujarat introduced an Android application, the SKY app, which provides daily system data to each farmer under SPFS. The app was designed with farmers’ linguistic needs in mind, offering a multilingual interface in English as well as vernacular languages such as Hindi and Gujarati. Primary data was collected primarily using unstructured conversations, passive observations, and notes during our interactions with the farmers. All participants in our study were males, reflecting the local agricultural and sociocultural context where land ownership and farm-level decision making are overwhelmingly male-dominated. A representative image of a SIP system can be seen in Figure \ref{fig:field}.

\subsection{Qualitative Analysis}
The lead author collected data through unstructured conversations, note-taking, and passive observation. One of the co-authors was already familiar with the farmers and their participation in the SPFS scheme and SPaRC pilot from the start, and conducted occasional site visits as needed. The lead/first author's engagement included field visits and phone conversations over more than two years. Given the longitudinal nature of the study, questions evolved, focusing primarily on farmers’ experiences with solar installation management and maintenance. We employed an "immersion/crystallization" approach, iteratively reflecting on our observations and farmer inputs to shape the findings \cite{crabtree2023doing}. Prolonged engagement with participants enhanced the credibility of the data.

%% file: 3.Findings.tex
\section{Findings}
In the sub-sections below, we detail the findings from the qualitative inputs from the farmers.
\subsection{Farmer's Perception of the Solar Schemes}
The SKY Android application provided farmers in C2 with system information such as current status, run hours, energy generation, and consumption, accessible via personal user IDs and passwords. It also included a contact window for reporting service disruptions. While farmers found these features useful for remotely monitoring their systems, they noted limitations, including the inability to connect with neighboring farmers or access maintenance contacts such as wiremen or software developers when needed. A farmer from C2 notes, 

\textit{"...application has contact for service, they help, but sometimes they are not available..."}.

\begin{figure}[htbp]
    \centering
    \includegraphics[width=1\linewidth]{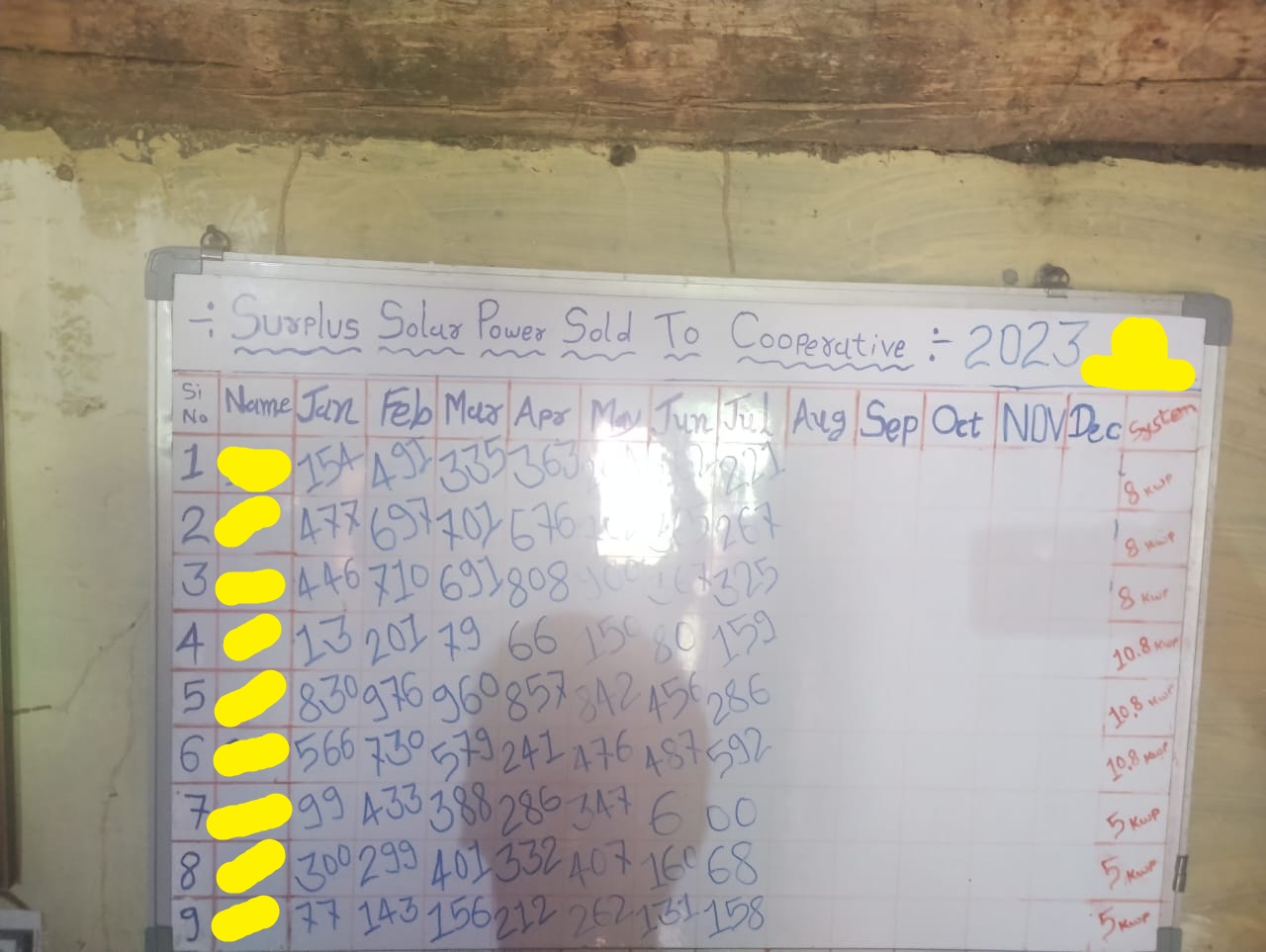}
    \caption{Energy data maintained physically by farmers in Cohort 1}
    \label{fig:offline}
    \Description{A photograph of a white dry-erase board contains a handwritten table in blue ink with the title "Surplus Solar Power Sold To Cooperative - 2023" at the top. The table has columns headed "Sl No", "Name", "Jan", "Feb", "Mar", "Apr", "May", "Jun", "Jul", "Aug", "Sep", "Oct", "NOV", "Dec", and "System" (indicating solar capacity). There are nine rows of numerical data listed under the months from January to July for different entries, with some data obscured by a light reflection in the center. The "Name" column and a small section in the top right are covered by yellow highlights so as to retain privacy. The "System" column lists values like "8 KWP", "10.8 KWP", and "5 KWP".}
\end{figure}

Farmers in C1 used an offline method to toll their energy exports and keep track of their generation, as can be seen in Figure \ref{fig:offline}. However, in C2, one proactive farmer created a WhatsApp group for sharing information and learning about solar installations. Farmers from both cohorts acknowledged the financial benefits of the solar schemes, but C2 farmers highlighted the need for more clarity in the SKY scheme financing. One farmer from C2 also highlighted the presence of groundwater markets, which might offset the grid evacuation incentives. A key point echoed by both C1 and C2 was the persistent maintenance challenges, with some farmers even resorting to legal action to resolve service issues. The next subsection highlights these challenges and an innovative solution devised by the C2 farmers to address some of these challenges.

\subsection{Overcoming Operational Challenges through Collective Action}
Farmers reported that SIPs installed far from their homes made timely maintenance difficult, and getting energy data from the field was tedious when the application would not function. As one farmer from C2 noted, 

\textit{"...if I did not have the application, it would be a terrible loss because we wouldn't know anything about the generation...If I am at my home or a different town, city, or village, I can open the application and see if my pump is running or stopped..."}

Both cohorts faced issues such as transformer faults, wire damage by insects, and dust accumulation on panels, and more, all of which reduced energy generation and directly affected earnings. To address this, the WhatsApp group became a platform for sharing best practices for cleaning and maintaining the systems in C2. Farmers also struggled to assess whether their panels were performing as expected, since the SKY app only showed individual status with no regional comparison. To overcome this, some farmers in C2 began sharing daily generation data within the WhatsApp group, comparing outputs with peers. Because all panels had the same capacity, deviations helped identify issues such as line losses, dust, shading, or other faults. As pointed out by a farmer, 

\textit{"...Sometimes the application shows zero for me but high for others...sometimes it's high for me but low for others, there must be some reason behind it, maybe dust accumulation, wire fault... "}.

Initially hesitant, most farmers gradually gained confidence in sharing data as progressive members led by example, allowing the group to collaboratively monitor system performance. 

%% file: 4.Discussions.tex
\section{Discussions \&  Future Work}
This section discusses how farmers appropriated WhatsApp to bridge the gaps in formal monitoring applications, transforming solitary maintenance tasks into collaborative community practices. We discuss how these grassroots digital infrastructures not only sustain decentralized solar networks but also offer pathways for future implementations to function as distributed climate-sensing grids through participatory governance and financial incentives.

\subsection{Digital Collaboration for Solar Maintenance}
While SIPs are generally low-maintenance, the panels require regular cleaning, as dust accumulation reduces energy output, affecting both water discharge and energy export to the grid \cite{https://doi.org/10.1002/wene.325}. Farmers in Southern Gujarat, under the SKY scheme, faced similar challenges in monitoring scattered SIPs installed far from their homes, which increased risks of theft and vandalism \cite{Patel19052025}. \citet{Patel19052025} also highlights opportunities for cooperatives to support maintenance. In our study, Farmers in C1 did not have access to the SKY application as they were the initial pilot on which SKY was modeled. They tallied their generated units physically by noting down values from electric meters every month, as regular visits to the panels were not convenient. As a result, the idea to use this data for detecting anomalies did not naturally evolve. On the other hand, farmers from C2 creatively used WhatsApp groups to collaboratively manage SIPs without relying solely on top-down approaches, by sharing data with each other daily. While the SKY Android app provided energy data and contact support, it lacked interactive peer-to-peer features. To fill this gap, farmers shared daily generation data and maintenance practices within the group, echoing prior studies showing that farmers often develop their own data collection methods \cite{10.1145/3637416}. This grassroots digital participation complements official monitoring systems, reduces utility burdens, and enables communities to collaboratively detect anomalies and troubleshoot issues. The social interactions, sharing knowledge, and comparing performance, were as crucial as the technical infrastructure itself. These findings highlight the role of informal digital platforms in sustaining decentralized solar networks through community-driven support.

\subsection{Implications for Decentralized Solar Implementation and Sustainability}
Our findings suggest that incentivizing farmers to evacuate excess energy to the grid is favored by the farmers, as indicated by their practice of appropriating WhatsApp for maintenance and management, which is outside the formal system set up by the government. Unlike centralized or large-scale solar farms, which face land acquisition, environmental, and management constraints \cite{doi:10.1073/pnas.1517656112, SAHOO2016927}, locally managed PV networks offer additional socio-environmental value. They transform irrigation infrastructure into dynamic, community-driven systems for energy performance, embedding farmers directly in decision-making processes. However, locally managed individual grid-connected solar pumps also face challenges that include grid-instability arising from reverse power injection; logistical, operational, and maintenance challenges for solar developers, and more \cite{Promotin35:online}. Additionally, as highlighted by prior work, the effect of groundwater markets on grid evacuation incentives needs further investigation \cite{Balasubramanya_2022}. These offer an opportunity to explore the potential of community-managed solar networks to deliver benefits beyond irrigation, in order to mitigate or compensate for their limitations and ensure their viability. Prior studies show that energy data from photovoltaic systems can be used to predict agro-climate variables \cite{7555541, su14074078}. By incentivizing farmers to share real-time solar generation data, decentralized solar systems could function as distributed climate-sensing grids, complementing broader citizen science and resilience initiatives \cite{https://doi.org/10.1002/ese3.162, kara2016realtimeestimationsolargeneration, delCanizo2021_GenerationSolar, Ginoya2021_BuiltToLast}. Integrating financial rewards with user-friendly digital platforms can also reduce operational and maintenance burdens often faced by electricity utilities and distribution companies, while promoting active farmer engagement. Furthermore, anonymized shared energy data could be potentially monetized, providing economic incentives that enhance the feasibility and appeal of decentralized renewable energy initiatives to all stakeholders, including farmers, distribution companies, service providers, and more. Future work should look into the feasibility of such capabilities. Overall, aligning technological design, digital tools, financial incentives, and policy support can strengthen both the sustainability and social impact of decentralized solar-powered agriculture.

%% file: 5.Conclusions.tex
\section{Conclusions}
This study highlights that the sustainability of decentralized energy systems in the Global South relies not just on robust engineering but also on effective management through social and digital fabrics that support them. We documented how farmers in rural Gujarat, faced with the challenges of maintaining remote solar irrigation pumps, tabulate energy data both physically and digitally by appropriating WhatsApp to construct an informal digital infrastructure for collective sensemaking. By sharing generation data and troubleshooting tips, they transformed a solitary monitoring task into a collaborative community practice, filling the gaps left by the formal application. Our findings suggest that digital participation is as crucial as the technical infrastructure itself. The grassroots digital innovation observed here offers a blueprint for future HCI interventions in agriculture. Rather than designing tools that view farmers as isolated rational actors, we argue for systems that support peer comparison, contextual interpretation, and community-driven maintenance. Moving forward, there is a significant opportunity to bridge these informal practices with formal systems and for exploration of solar networks for agro-climatic sensing. By incentivizing the sharing of solar generation data, decentralized agricultural networks could potentially evolve into distributed climate-sensing grids and support citizen science initiatives. Ultimately, this work frames farmer-managed solar irrigation as a socio-technical platform, where local collaboration and everyday digital tools are fundamental to achieving long-term energy resilience and sustainability.

%% file: references.bib
@article{AHMED2025100848,
title = {Advancing agriculture through IoT, Big Data, and AI: A review of smart technologies enabling sustainability},
journal = {Smart Agricultural Technology},
volume = {10},
pages = {100848},
year = {2025},
issn = {2772-3755},
doi = {https://doi.org/10.1016/j.atech.2025.100848},
url = {https://www.sciencedirect.com/science/article/pii/S2772375525000814},
author = {Nurzaman Ahmed and Nadia Shakoor}
}

@article{jsan13040039,
AUTHOR = {Fuentes-Peñailillo, Fernando and Gutter, Karen and Vega, Ricardo and Silva, Gilda Carrasco},
TITLE = {Transformative Technologies in Digital Agriculture: Leveraging Internet of Things, Remote Sensing, and Artificial Intelligence for Smart Crop Management},
JOURNAL = {Journal of Sensor and Actuator Networks},
VOLUME = {13},
YEAR = {2024},
NUMBER = {4},
ARTICLE-NUMBER = {39},
URL = {https://www.mdpi.com/2224-2708/13/4/39},
ISSN = {2224-2708},
DOI = {10.3390/jsan13040039},
pages = {1--26}
}

@article{PARRALOPEZ2024109412,
title = {Integrating digital technologies in agriculture for climate change adaptation and mitigation: State of the art and future perspectives},
journal = {Computers and Electronics in Agriculture},
volume = {226},
pages = {109412},
year = {2024},
issn = {0168-1699},
doi = {https://doi.org/10.1016/j.compag.2024.109412},
url = {https://www.sciencedirect.com/science/article/pii/S0168169924008032},
author = {Carlos Parra-López and Saker {Ben Abdallah} and Guillermo Garcia-Garcia and Abdo Hassoun and Pedro Sánchez-Zamora and Hana Trollman and Sandeep Jagtap and Carmen Carmona-Torres}
}

@article{Ahmadi2025,
  author    = {Ahmadi, Senour and Amendolagine, Vito and LaSala, Piermichele},
  title     = {Unpacking the impacts of digitalization of knowledge transfer in agri-food sector, through sociotechnical systems theory: a systematic literature review},
  journal   = {Agricultural Economics},
  volume    = {13},
  pages     = {51},
  year      = {2025},
  doi       = {10.1186/s40100-025-00393-3},
  url       = {https://doi.org/10.1186/s40100-025-00393-3},
}

@online{IIITAres22:online,
author = {Rajeev Mani},
  title = {IIIT-A researchers develop AI tech for real time crop disease detection in Indian farms | The Times of India},
  url = "https://timesofindia.indiatimes.com/city/allahabad/iiit-a-researchers-develop-ai-tech-for-real-time-crop-disease-detection-in-indian-farms/articleshow/123370013.cms",
month = {August},
year = {2025},
  note = "Online; Accessed 25th February 2026"
}

@online{AIforagr79:online,
author = {World Economic Forum},
  title = {AI for agriculture: How Indian farmers are harnessing emerging technologies to sustainably increase productivity},
  url = "https://www.weforum.org/impact/ai-for-agriculture-in-india/",
  month = {1},
  year = {2024},
  note = "Online; Accessed 25th February 2026"
}

@online{Cabinetn66:online,
author = {The Times of India},
  title = {Cabinet nod for agri AI policy, to allot 500cr for first 3 years},
  url = "https://timesofindia.indiatimes.com/city/mumbai/cabinet-nod-for-agri-ai-policy-to-allot-500cr-for-first-3-years/articleshow/121915447.cms",
  month = {June},
  year = {2025},
  note = "Online; Accessed 25th February 2026"
}

@online{Jeevanandam2024,
  author       = {Jeevanandam, Nivash},
  title        = {AI in agriculture in 2025: Transforming Indian farms for a sustainable future},
  howpublished = {IndiaAI (online article)},
  month        = {December},
  year         = {2024},
  url          = {https://indiaai.gov.in/article/ai-in-agriculture-in-2025-transforming-indian-farms-for-a-sustainable-future},
  note = {Online; Accessed 25th February 2026}
}

@article{Bhat2025,
  author       = {Bhat, Irshad Ahmad and Ansarullah, Syed Immamul and Ahmad, Fearoz and Amir, Sheikh and Sidana, Sagar and Sinha, Anurag and Khalid, Saifullah and Yazdani, Ghulam},
  title        = {Leveraging Artificial Intelligence in Agribusiness: A Structured Review of Strategic Management Practices and Future Prospects},
  journal      = {Discover Sustainability},
  volume       = {6},
  number       = {565},
  pages        = {},
  year         = {2025},
  doi          = {10.1007/s43621-025-01260-3},
  url          = {https://doi.org/10.1007/s43621-025-01260-3}
}

@article{Odume2024,
  author    = {Odume, Blessing Winifred},
  title     = {Artificial Intelligence in Agriculture: Application in Developing Countries},
  journal   = {Journal of Agricultural Science},
  volume    = {16},
  number    = {12},
  pages     = {60--67},  
  year      = {2024},
  doi       = {10.5539/jas.v16n12p60},
  url       = {https://ccsenet.org/journal/index.php/jas/article/view/0/50910},
}

@inproceedings{10.1145/3313831.3376364,
author = {Pschetz, Larissa and Dixon, Billy and Pothong, Kruakae and Bailey, Arlene and Glean, Allister and Soares, Luis Louren\c{c}o and Enright, Jessica A.},
title = {Designing Distributed Ledger Technologies for Social Change: The Case of CariCrop},
year = {2020},
isbn = {9781450367080},
publisher = {Association for Computing Machinery},
address = {New York, NY, USA},
url = {https://doi.org/10.1145/3313831.3376364},
doi = {10.1145/3313831.3376364},
booktitle = {Proceedings of the 2020 CHI Conference on Human Factors in Computing Systems},
pages = {1–12},
numpages = {12},
location = {Honolulu, HI, USA},
series = {CHI '20}
}

@article{10.1145/3274418,
author = {Leshed, Gilly and Rosca, Masha and Huang, Michael and Mansbach, Liza and Zhu, Yicheng and Hern\'{a}ndez-Aguilera, Juan Nicol\'{a}s},
title = {CalcuCaf\'{e}: Designing for Collaboration Among Coffee Farmers to Calculate Costs of Production},
year = {2018},
issue_date = {November 2018},
publisher = {Association for Computing Machinery},
address = {New York, NY, USA},
volume = {2},
number = {CSCW},
url = {https://doi.org/10.1145/3274418},
doi = {10.1145/3274418},
journal = {Proc. ACM Hum.-Comput. Interact.},
month = nov,
articleno = {149},
numpages = {26}
}

@article{10.1145/3432932,
author = {Hanrahan, Benjamin V. and Maitland, Carleen and Brown, Timothy and Chen, Anita and Kagame, Fraterne and Birir, Beatrice},
title = {Agency and Extraction in Emerging Industrial Drone Applications: Imaginaries of Rwandan Farm Workers and Community Members},
year = {2021},
issue_date = {December 2020},
publisher = {Association for Computing Machinery},
address = {New York, NY, USA},
volume = {4},
number = {CSCW3},
url = {https://doi.org/10.1145/3432932},
doi = {10.1145/3432932},
journal = {Proc. ACM Hum.-Comput. Interact.},
month = jan,
articleno = {233},
numpages = {21}
}

@inproceedings{10.1145/3613904.3642263,
author = {Doggett, Olivia and Ratto, Matt and Chandra, Priyank},
title = {Migrant Farmworkers' Experiences of Agricultural Technologies: Implications for Worker Sociality and Desired Change},
year = {2024},
isbn = {9798400703300},
publisher = {Association for Computing Machinery},
address = {New York, NY, USA},
url = {https://doi.org/10.1145/3613904.3642263},
doi = {10.1145/3613904.3642263},
booktitle = {Proceedings of the 2024 CHI Conference on Human Factors in Computing Systems},
articleno = {563},
numpages = {23},
location = {Honolulu, HI, USA},
series = {CHI '24}
}

@inproceedings{10.1145/3613904.3642099,
author = {Raghunath, Ananditha and Metzger, Alexander Le and Easton, Hans and Liu, Xunmei and Wang, Fanchong and Wang, Yunqi and Zhao, Yunwei and Mpogole, Hosea and Anderson, Richard},
title = {eKichabi v2: Designing and Scaling a Dual-Platform Agricultural Technology in Rural Tanzania},
year = {2024},
isbn = {9798400703300},
publisher = {Association for Computing Machinery},
address = {New York, NY, USA},
url = {https://doi.org/10.1145/3613904.3642099},
doi = {10.1145/3613904.3642099},
booktitle = {Proceedings of the 2024 CHI Conference on Human Factors in Computing Systems},
articleno = {503},
numpages = {16},
location = {Honolulu, HI, USA},
series = {CHI '24}
}

@inproceedings{10.1145/3706598.3713643,
author = {Chaudhary, Akash and Escobar, Stefany Arevalo and Zayas, Dulce and Su, Norman Makoto},
title = {Normalizing Grit: The Futility of Personal Informatics for Farm Workers and Climate Change},
year = {2025},
isbn = {9798400713941},
publisher = {Association for Computing Machinery},
address = {New York, NY, USA},
url = {https://doi.org/10.1145/3706598.3713643},
doi = {10.1145/3706598.3713643},
booktitle = {Proceedings of the 2025 CHI Conference on Human Factors in Computing Systems},
articleno = {949},
numpages = {17},
location = {
},
series = {CHI '25}
}

@article{10.1145/3610222,
author = {Medarametla, Lokesh and Ahmed, Saquib and Dey, Sanorita},
title = {Why should I bother to break the norm?: Exploring the Prospects of Adopting Technology-Driven Solutions by Indian Shrimp Farmers},
year = {2023},
issue_date = {October 2023},
publisher = {Association for Computing Machinery},
address = {New York, NY, USA},
volume = {7},
number = {CSCW2},
url = {https://doi.org/10.1145/3610222},
doi = {10.1145/3610222},
journal = {Proc. ACM Hum.-Comput. Interact.},
month = oct,
articleno = {373},
numpages = {25}
}

@techreport{IEA2024Renewables,
  author       = {{International Energy Agency (IEA)}},
  title        = {Renewables 2024: Analysis and forecasts to 2030},
  institution  = {International Energy Agency},
  address      = {Paris},
  year         = {2024},
  month        = {October},
  note         = {Licence: CC BY 4.0; Online; Accessed 25th February 2026},
  url          = {https://www.iea.org/reports/renewables-2024},
}

@techreport{IRENA2024RenewableStats,
  author       = {{International Renewable Energy Agency (IRENA)}},
  title        = {Renewable Energy Statistics 2024},
  institution  = {International Renewable Energy Agency},
  address      = {Abu Dhabi},
  year         = {2024},
  month        = {July},
  url          = {https://www.irena.org/Publications/2024/Jul/Renewable-energy-statistics-2024 },
  note = {Online; Accessed 25th February 2026},
  isbn = {978-92-9260-614-5}
}

@online{PIB2024RenewableMilestone,
  author       = {{Press Information Bureau, Government of India}},
  title        = {India’s Renewable Energy Capacity Hits New Milestone},
  howpublished = {Press Information Bureau press release},
  institution  = {Ministry of New and Renewable Energy, Government of India},
  address      = {New Delhi},
  month        = {November},
  day          = {13},
  year         = {2024},
  url          = {https://www.pib.gov.in/PressReleasePage.aspx?PRID=2073038},
  note = {Online; Accessed 4th March 2026}
}

@online{MSEDCL_iSMART2025,
  author       = {{Maharashtra State Electricity Distribution Company Limited (MSEDCL)}},
  title        = {i-SMART: MSEDCL’s Integrated Solar Mobile and Rooftop Platform},
  howpublished = {Official MSEDCL web portal},
  institution  = {MSEDCL},
  address      = {Mumbai, India},
  year         = {[2025]},
  note         = "Online; Accessed 25th February 2026",
  url          = {https://www.mahadiscom.in/ismart/},
}

@online{Deshpande2025MSEDCL,
  author       = {Deshpande, Viraj},
  title        = {MSEDCL Chief Honoured With Energy Leadership Award 2025},
  journal      = {Times of India},
  year         = {2025},
  month        = {September},
  day          = {9},
  note         = "Online; Accessed 25th February 2026",
  url          = {https://timesofindia.indiatimes.com/city/nagpur/msedcl-chief-honoured-withenergy-leadership-award-2025/articleshow/123788020.cms},
}

@online{GujaratEPD_AgricultureSchemes2025,
  author       = {{Energy \& Petrochemicals Department, Government of Gujarat}},
  title        = {Agriculture-Related Schemes},
  howpublished = {Official government webpage (Blog Details Page 31)},
  institution  = {Energy \& Petrochemicals Department, Government of Gujarat},
  address      = {Gandhinagar, Gujarat, India},
  year         = {[2025]},
  note         = "Online; Accessed 25th February 2026",
  url          = {https://guj-epd.gujarat.gov.in/Home/BlogDetailsPage/31},
}

@online{GujaratSolarPolicy2021,
  author       = {Energy \& Petrochemicals Department, Government of Gujarat},
  title        = {Gujarat Solar Power Policy 2021},
  howpublished = {Official policy webpage},
  institution  = {Energy \& Petrochemicals Department, Government of Gujarat},
  address      = {Gandhinagar, Gujarat, India},
  year         = {[2021]},
  url          = {https://guj-epd.gujarat.gov.in/Home/gujaratsolarpowerpolicy},
  note         = "Online; Accessed 25th February 2026"
  
}

@online{IPAG_DRE_2024,
  author       = {Syed Munir Khasru and Riad Meddeb and Gauri Singh and Vibha Dhawan and Shailly Kedia and Sarwat Chowdhury},
  title        = {Decentralized Renewable Energy (DRE) Systems: A Pathway to Just Energy Transitions in Vulnerable Communities},
  howpublished = {IPAG Policy Brief},
  institution  = {Institute for Policy, Advocacy, and Governance (IPAG)},
  address      = {Global},
  year         = {2024},
  month        = {October},
  day          = {7},
  note         = {Online; Accessed 25th February 2026},
  url          = {https://www.ipag.org/policy/decentralized-renewable-energy-dre-systems-a-pathway-to-just-energy-transitions-in-vulnerable-communities/ },
}

@online{TOI2025_RajasthanRenewableEnergy,
  author       = {{Times of India}},
  title        = {CPUs delay transmission projects, 25\% renewable energy goes waste},
  journal      = {The Times of India},
  year         = {2025},
  month        = {September},
  day          = {9},
  url          = {https://timesofindia.indiatimes.com/city/jaipur/cpus-delay-transmission-projects-25-renewable-energy-goes-waste/articleshow/123773065.cms},
  note         = {Online; Accessed 25th February 2026},
}

@online{Raizada2025_IndiaEnergyTransition,
  author       = {Akul Raizada},
  title        = {Unlocking India’s Energy Transition: Addressing Grid Flexibility Challenges and Solutions},
  howpublished = {Policy Memo},
  institution  = {Institut français des relations internationales (Ifri)},
  address      = {Paris, France},
  year         = {2025},
  month        = {February},
  day          = {20},
  url          = {https://www.ifri.org/en/memos/unlocking-indias-energy-transition-addressing-grid-flexibility-challenges-and-solutions },
  note         = {Online; Accessed 25th February 2026},
}

@inproceedings{10.1145/3706598.3714250,
author = {Singh, Anubha and Garcia, Patricia and Chandra, Priyank},
title = {What's in a Place? On Platformization of Traditional Agricultural Marketplaces},
year = {2025},
isbn = {9798400713941},
publisher = {Association for Computing Machinery},
address = {New York, NY, USA},
url = {https://doi.org/10.1145/3706598.3714250},
doi = {10.1145/3706598.3714250},
booktitle = {Proceedings of the 2025 CHI Conference on Human Factors in Computing Systems},
articleno = {952},
numpages = {16},
location = {
},
series = {CHI '25}
}

@inproceedings{10.1145/3678884.3682046,
author = {Porf\'{\i}rio, Rui Pedro},
title = {Studying the Impact of Explainable AI in Digital Agriculture Solutions},
year = {2024},
isbn = {9798400711145},
publisher = {Association for Computing Machinery},
address = {New York, NY, USA},
url = {https://doi.org/10.1145/3678884.3682046},
doi = {10.1145/3678884.3682046},
booktitle = {Companion Publication of the 2024 Conference on Computer-Supported Cooperative Work and Social Computing},
pages = {43–46},
numpages = {4},
location = {San Jose, Costa Rica},
series = {CSCW Companion '24}
}

@article{10.1145/3359136,
author = {Norton, Juliet and Penzenstadler, Birgit and Tomlinson, Bill},
title = {Implications of Grassroots Sustainable Agriculture Community Values on the Design of Information Systems},
year = {2019},
issue_date = {November 2019},
publisher = {Association for Computing Machinery},
address = {New York, NY, USA},
volume = {3},
number = {CSCW},
url = {https://doi.org/10.1145/3359136},
doi = {10.1145/3359136},
journal = {Proc. ACM Hum.-Comput. Interact.},
month = nov,
articleno = {34},
numpages = {22}
}

@Article{agronomy14122791,
AUTHOR = {Espitia, John Javier and Velázquez, Fabián Andrés and Rodriguez, Jader and Gomez, Luisa and Baeza, Esteban and Aguilar-Rodríguez, Cruz Ernesto and Flores-Velazquez, Jorge and Villagran, Edwin},
TITLE = {Solar Energy Applications in Protected Agriculture: A Technical and Bibliometric Review of Greenhouse Systems and Solar Technologies},
JOURNAL = {Agronomy},
VOLUME = {14},
YEAR = {2024},
NUMBER = {12},
ARTICLE-NUMBER = {2791},
URL = {https://www.mdpi.com/2073-4395/14/12/2791},
ISSN = {2073-4395},
DOI = {10.3390/agronomy14122791},
pages = {1--31}
}

@online{DOE_Agrivoltaics,
  author       = {{U.S. Department of Energy}},
  title        = {Agrivoltaics: Solar and Agriculture Co-Location},
  url = {https://www.energy.gov/eere/solar/agrivoltaics-solar-and-agriculture-co-location},
  year         = {2023},
  note         = {Online; Accessed 25th February 2026},
}

@online{Gilchrist2024,
  author       = {Duncan Gilchrist and Joanna Kulesza},
  title        = {Agrivoltaic Farms Grow Both Solar Power and Food in Colorado},
  howpublished = {\url{https://www.nature.org/en-us/magazine/magazine-articles/agrivoltaic-solar-farm-grows-produce/}},
  year         = {2024},
  month        = may,
  url = {https://www.nature.org/en-us/magazine/magazine-articles/agrivoltaic-solar-farm-grows-produce/},
  note         = {Online; Accessed 25th February 2026}
}

@online{Macknick2019_Agrivoltaics_NREL,
  author       = {National Laboratory of the Rockies (NLR)},
  title        = {Benefits of Agrivoltaics Across the Food-Energy-Water Nexus},
  howpublished = {\url{https://www.nlr.gov/grid/news/program/2019/benefits-of-agrivoltaics-across-the-food-energy-water-nexus}},
  year         = {2019},
  month        = {sep},
  note         = {Online; Accessed 25th February 2026}
}

@techreport{UNEPDHI2024_WEFE_Nexus,
  author       = {{UNEP-DHI Centre on Water and Environment} and {UNEP Copenhagen Climate Centre}},
  title        = {Working with the Water-Energy-Food-Ecosystems (WEFE) Nexus: Policy Brief},
  url = {https://unepdhi.org/working-with-the-water-energy-food-ecosystems-wefe-nexus/},
  year         = {2024},
  institution  = {UNEP-DHI Centre and Copenhagen Climate Centre},
  note         = {Online; Accessed 25th February 2026}
}

@online{UNECE_WEFE_Nexus,
  author       = {{United Nations Economic Commission for Europe (UNECE)}},
  title        = {Water-Food-Energy-Ecosystem Nexus: Areas of Work under the Water Convention},
  url = {https://unece.org/environment-policy/water/areas-work-convention/water-food-energy-ecosystem-nexus},
  year         = {2025},
  note         = {Online; Accessed 25th February 2026},
}

@online{Shah2024_BiharSolarPumps,
  author       = {Shah, Tushaar and Singh, Meenakshi and Patidar, Naveen},
  title        = {Solar Pumps are Key to Bihar’s Rural Electrification Strategy},
  howpublished = {\url{https://www.theindiaforum.in/sites/default/files/article_pdf/2024/11/22/1681-1732264265.pdf}},
  month        = {nov},
  year         = {2024},
  url = {https://www.theindiaforum.in/sites/default/files/article_pdf/2024/11/22/1681-1732264265.pdf},
  note         = {Online; Accessed 25th February 2026},
}

@online{Shah2016_SolarCrop,
  author       = {Shah, Tushaar and Durga, Neha and Verma, Shilp and Rathod, Rahul},
  title        = {Solar Power as Remunerative Crop},
  url = {https://hdl.handle.net/10568/80877},
  howpublished = {Research Highlight},
  publisher    = {International Water Management Institute - Tata Water Policy Program},
  year         = {2016},
  note = {Online; Accessed 25th February 2026}
}

@article{Durga2021_SuryaRaithaEPW,
  author = {Durga, Neha and Shah, Tushaar and Verma, Shilp and Manjunatha A. V.},
  title = {Karnataka’s ‘Surya Raitha’ Experiment: Lessons for PM–KUSUM},
  url = "https://www.epw.in/journal/2021/48/special-articles/karnatakas-surya-raitha-experiment.html",
  volume = {56},
  number = {48},
  pages = {55--60},
  year = {2021},
  journal={Economic and Political Weekly},
  note = "Online; Accessed 25th February 2026"
}

@online{TOI_HaryanaSolarGrowth2025,
  author       = {{Times of India}},
  title        = {Green Energy Boost: Haryana’s Solar Power Generation Sees Growth},
  howpublished = {News aricle},
  url = {https://timesofindia.indiatimes.com/city/gurgaon/green-energy-boost-haryanas-solar-power-generation-sees-growth/articleshow/121146887.cms},
  year         = {2025},
  month        = {aug},
  note         = {Online; Accessed 25th February 2026},
}

@online{TOI_AP_FarmSolarisation2025,
  author       = {{Times of India}},
  title        = {State Plans Full Solarisation of Farm Power in One Year},
  howpublished = {News Article},
  url          = {https://timesofindia.indiatimes.com/city/vijayawada/state-plans-full-solarisation-of-farm-power-in-one-year/articleshow/123487312.cms},
  year         = {2025},
  month        = {aug},
  note         = {Online; Accessed 4th March 2026},
}

@article{ChoquetteLevy2025,
  author       = {Nicolas Choquette-Levy and Dirgha Ghimire and Michael Oppenheimer and Rajendra Ghimire and Dil Ck},
  title        = {Retrenchment under climate-driven risks in subsistence farming communities},
  journal      = {Population and Environment},
  volume       = {47},
  number       = {22},
  pages        = {35},
  year         = {2025},
  doi          = {10.1007/s11111-025-00493-8},
  url          = {https://doi.org/10.1007/s11111-025-00493-8}
}

@article{Mitter2024,
  author       = {Hermine Mitter and Kathrin Obermeier and Erwin Schmid},
  title        = {Exploring smallholder farmers’ climate change adaptation intentions in Tiruchirappalli District, South India},
  journal      = {Agriculture and Human Values},
  volume       = {41},
  number       = {},
  pages        = {1019--1035},
  year         = {2024},
  doi          = {10.1007/s10460-023-10528-1},
  url          = {https://link.springer.com/article/10.1007/s10460-023-10528-1},
}

@article{Jatav2024,
  author       = {Surendra Singh Jatav},
  title        = {Farmers’ perception of climate change and livelihood vulnerability: a comparative study of Bundelkhand and Central regions of Uttar Pradesh, India},
  journal      = {Discover Sustainability},
  volume       = {5},
  number      = {11},
  year         = {2024},
  doi          = {10.1007/s43621-024-00193-7},
  url          = {https://link.springer.com/article/10.1007/s43621-024-00193-7},
  pages = {}
}

@article{Thein2025,
title = {Assessing livelihood vulnerability to climate change in rural India},
journal = {World Development Sustainability},
volume = {7},
pages = {100249},
year = {2025},
issn = {2772-655X},
doi = {https://doi.org/10.1016/j.wds.2025.100249},
url = {https://www.sciencedirect.com/science/article/pii/S2772655X25000473},
author = {Kyawt Yin Min Thein and Vivek Kumar and Vijayaraghavan M Chariar and Takuji W. Tsusaka}
}

@article{TheCasef55:online,
author = {Tushaar Shah and  Arnab Paul Choudhury and Rahul Rathod and Gyan P. Rai and Shilp Verma},
title = {The Case for Grid-connected Solar Irrigation Pumps | Economic and Political Weekly},
volume = {60},
number = {11},
journal={Economic and Political Weekly},
month = {June},
year = {2025},
doi = {10.71279/epw.v60i11.38697},
URL = {  
        https://doi.org/10.71279/epw.v60i11.38697
    },
pages = {52--59}
}

@online{GridConn12:online,
author = {Deepak Varshney and Aditi Mukherji and Kriti Sharma and Alok Sikka},
  title = {Pre-print: Grid-Connected Solar Irrigation Pumps for Farmers: An Evaluation of Gujarat's Surya Shakti Kisan Yojana (Sky)},
  url = {https://papers.ssrn.com/sol3/papers.cfm?abstract_id=5041875},
  doi = {https://dx.doi.org/10.2139/ssrn.5041875},
  month = {December},
  year = {2024},
  note = "Online; Accessed 25th February 2026"
}

@online{UGVCL87:online,
author = {Uttar Gujarat Vij Company Limited},
  title = {Surya Shakti Kisan Yojana},
  url = "https://www.ugvcl.com/SuryaShaktiKishanyojana",
  year = {[2018]},
  note = "Online; Accessed 25th February 2026"
}

@book{crabtree2023doing,
  title={Doing qualitative research},
  author={Crabtree, Benjamin F and Miller, William L},
  year={2023},
  publisher={Sage Publications},
  address={Thousand Oaks, CA},
  edition={3}
}

@online{GujaratH66:online,
author = {Deryck O. Lodrick, Devavrat Nanubhai Pathak},
  title = {Gujarat | History, Map, Population, \& Facts | Britannica},
  year = {2026},
  url = "https://www.britannica.com/place/Gujarat",
  note = "Online; Accessed 25th February 2026"
}

@article{https://doi.org/10.1002/wene.325,
author = {Agrawal, Shalu and Jain, Abhishek},
title = {Sustainable deployment of solar irrigation pumps: Key determinants and strategies},
journal = {WIREs Energy and Environment},
volume = {8},
number = {2},
pages = {e325},
doi = {https://doi.org/10.1002/wene.325},
url = {https://wires.onlinelibrary.wiley.com/doi/abs/10.1002/wene.325},
year = {2019}
}

@article{Patel19052025,
author = {Praharsh M. Patel and Christopher A. Scott and Daniel A. Brent},
title = {Four birds with one stone? opportunities and challenges in adopting solar irrigation for a sustainable water-energy-food nexus with carbon credits},
journal = {Water International},
volume = {50},
number = {3-4},
pages = {235--258},
year = {2025},
publisher = {Routledge},
doi = {10.1080/02508060.2024.2345494},
URL = { 
        https://doi.org/10.1080/02508060.2024.2345494
},

}

@article{SAHOO2016927,
title = {Renewable and sustainable energy reviews solar photovoltaic energy progress in India: A review},
journal = {Renewable and Sustainable Energy Reviews},
volume = {59},
pages = {927-939},
year = {2016},
issn = {1364-0321},
doi = {https://doi.org/10.1016/j.rser.2016.01.049},
url = {https://www.sciencedirect.com/science/article/pii/S1364032116000794},
author = {Sarat Kumar Sahoo}
}
